\documentclass[reprint,aps]{revtex4-2}
\usepackage{amsfonts}
\usepackage{eurosym}
\usepackage{amssymb}
\usepackage{amsmath}
\usepackage{tabularx,epsfig,color,tabularx,epstopdf}
\usepackage{graphicx}
\usepackage{dcolumn}
\usepackage{bm}

\usepackage{epstopdf}

\begin{document}

\title{Exact vector Akhmediev breathers dominated by a linearly stable frequency}
\author{Wen-Rong Sun$^{1}$}
\author{Chong Liu$^{2}$}
\email{chongliu@nwu.edu.cn}
\author{Lei Wang$^{3}$}
\author{Fabio Baronio$^{4}$}
\affiliation{1. School of Mathematics and Physics, University of Science and Technology Beijing, Beijing 100083, China\\
2. School of Physics, Northwest University, Xi’an 710127, China\\
3. School of Mathematics and Physics, North China Electric Power University, Beijing 102206, China\\
4. Department of Information Engineering, University of Brescia, 25123 Brescia, Italy}

\begin{abstract}
In the scalar nonlinear Schr\"{o}dinger equation, an Akhmediev breather (AB) is dominated by a frequency that lies inside the modulation instability gain band. This exactly correspondence between instability and breathers is challenged in vector systems such as the Manakov system, where the gain spectrum splits into disconnected lobes separated by stable gaps. We analytically and numerically construct an AB that is generated by unstable modes but is spectrally dominated by a stable harmonic at its peak. 
Numerical simulations starting from a simple continuous‑wave background perturbed only by the unstable harmonics confirm that the stable 
component emerges spontaneously and becomes dominant without any
initial seed.
We identify the precise parameter window
in which this phenomenon occurs and show that this passive amplification
of the linearly stable component is driven by four-wave mixing, which
accounts for $96\%$ of the nonlinear forcing. 
Given the universality of the Manakov system across nonlinear physics, from
nonlinear optics to ultracold quantum gases, these results open an experimentally
accessible new perspective on breather dynamics, one in which linearly stable
frequencies can dominate.
\end{abstract}

\maketitle

Modulation instability (MI) is a fundamental process through which a continuous wave
breaks up into localized structures in nonlinear dispersive media~\cite{mi1,mi2}.
In the scalar nonlinear Schr\"{o}dinger equation (NLSE), the nonlinear stage of MI
is well understood: a weak modulation at a frequency within the unstable band
excites an Akhmediev breather (AB)~\cite{ab1,ab2}, an exact solution describing a single
growth-decay cycle. Higher-order MI~\cite{hab1,hab2,hab3}, involving multiple unstable spectral components,
can be captured by multi-AB solutions~\cite{hab4}. In such case, each elementary
breather expands a pair of initial sidebands into higher harmonics, and the resulting
wave field is dominated by those frequency components that lie within the MI gain bands.

Moving beyond the scalar NLSE to more general classes of coupled wave systems
reveals richer features of both the MI spectrum and the nonlinear
stage of the dynamics~\cite{appl1,appl2,appl3,nd1,nd2,nd3,nd4}. An integrable model of practical importance is the Manakov system~\cite{ma1}.
Its integrability, combined with its direct applicability to coupled
nonlinear waves in settings ranging from nonlinear optics~\cite{appl1, appl4,appl5,appl6,appl7,appl8, appl9, nd10,nd9,lc1,appl10} to quantum
gases~\cite{appl3} and hydrodynamics~\cite{appl2}, has made it a central platform for exploring
dynamics of vector nonlinear waves and MI.  One of the intriguing aspects of the Manakov MI spectrum is that the gain curve splits into disconnected lobes
separated by stable gaps---frequency intervals where linear stability analysis
predicts zero growth~\cite{appl8} .  In all previous works related to the AB of the Manakov system, however,
the breather structures that emerge are invariably associated with
frequency components that reside inside the unstable bands. The stable
gaps in Manakov system, although a well-documented feature of the linear MI spectrum,
have been regarded as dynamically inactive: no self-amplified
modulation can occur there, and consequently no AB is expected to form
at those frequencies. Note that an AB is a product of the nonlinear
stage of MI. In this stage, four-wave mixing can significantly expand the
wave-field spectrum far beyond the initial modulation sidebands~\cite{du1}. This
naturally raises the question: within the stable gap of the Manakov system,
where linear theory forbids any self-amplified modulation, can the nonlinear
stage, driven by four-wave mixing, produce an AB that is dominated by a frequency
component residing inside the stable gap?

In this Letter, we analytically and numerically show that a third-order AB,
constructed solely from unstable harmonics ($\omega$ and $3\omega$),
can develop a dominant spectral peak at the linearly stable frequency $2\omega$,
even though linear theory forbids any self-amplified modulation at that frequency.
Since ABs are generally viewed as nonlinear manifestations of MI, their dominant 
spectral content is expected to remain inside the MI gain bands. The exact solution we 
present violates this intuitive correspondence: although generated entirely by 
unstable modes, the breather's peak spectrum is controlled by a harmonic with 
strictly zero linear gain.
Numerical simulations starting from a simple  continuous‑wave (CW) background (and exact solution) perturbed only by the unstable harmonics confirm that the stable 
component emerges spontaneously and becomes dominant without any
initial seed. The mechanism is passive nonlinear driving: four-wave mixing between the
unstable modes continuously transfers energy to the stable harmonic,
eventually making it the leading component of the wave profile. It firmly establishes that the stable $2\omega$ component, rather than being bypassed as in the frequency-jumping scenario~\cite{appl8}, becomes the leading spectral feature through passive nonlinear driving.

In the focusing regime, the dimensionless Manakov system
\begin{equation}
i\partial_t\psi^{(j)}+\frac12\partial_x^2\psi^{(j)}+(|\psi^{(1)}|^2+|\psi^{(2)}|^2)\psi^{(j)}=0,\quad j=1,2,
\label{eq:Manakov}
\end{equation}
governs the evolution of two nonlinearly coupled wave components $\psi^{(1)}(x,t)$ and $\psi^{(2)}(x,t)$. The physical meaning of the independent variables $x$ and $t$  depends
on the specific physical context.

Equation~(\ref{eq:Manakov}) admits a CW background solution
\begin{equation}
\psi_0^{(j)}(x,t)=a e^{i(\beta_j x+\Omega t)},\qquad j=1,2,
\label{eq:CW}
\end{equation}
with amplitude $a$, relative wavenumbers $\beta_1=-\beta_2=\beta>0$, and common frequency $\Omega=2a^2-\beta^2/2$. The parameter $\beta$ quantifies the wavenumber offset between the two components and serves as the main control parameter in our analysis. 
On this background, a fundamental AB of the Manakov system 
is a one-parameter family of exact solutions parametrized by the modulation 
frequency $\omega$. Its explicit form involves the spectral parameter~\cite{appl8}
\begin{equation}
\chi_\pm(\omega) = 
\pm\sqrt{\beta^2-a^2+\frac{\omega^2}{4}-\sqrt{a^4+(\omega^2-4a^2)\beta^2}}-\frac{\omega}{2},
\label{eq:chi_omega1}
\end{equation}
which determines the growth rate $G = |\omega\operatorname{Im}\,\chi_\pm|$. 
A fundamental AB exists only when $\omega$ lies inside 
the MI gain band.  Throughout this work, we set the background amplitude 
to $a=1$ and the fundamental modulation frequency to $\omega=1$ without 
loss of generality.

To explore the possibility of creating an AB controlled by a linearly stable frequency, we focus on the parameter window where the first and third 
harmonics are modulationally unstable while the second harmonic remains strictly stable. 
We consider a weakly modulated wave of the form $\psi^{(j)}=\psi_0^{(j)}(1+u^{(j)})$, where $|u^{(j)}|\ll 1$ is a small complex perturbation. Expanding $u^{(j)}$ in a Fourier series with the fundamental frequency $\omega=1$,
\begin{equation}
u^{(j)}(x,t)=\sum_{n=-\infty}^\infty a_n^{(j)}(t)\,e^{i n\omega x},
\label{eq:Fourier}
\end{equation}
and retaining only the linear terms in Eq.~(\ref{eq:Manakov}), we obtain a closed system for the Fourier amplitudes of each harmonic $n$. Owing to the coupling between positive and negative frequency components through complex conjugation, it is convenient to introduce the 4‑component vector
\begin{equation}
\mathbf{V}_n(t)=\bigl(a_n^{(1)}(t),\; a_n^{(2)}(t),\; a_{-n}^{(1)*}(t),\; a_{-n}^{(2)*}(t)\bigr)^{\!\mathrm{T}},
\label{eq:Vvector}
\end{equation}
whose dynamics obeys $i\,d\mathbf{V}_n/dt = \mathbf{M}_n\mathbf{V}_n$, with the $4\times4$ stability matrix
$(\mathbf{M}_n)_{11} = \frac{n^2}{2}+n\beta-1$, $(\mathbf{M}_n)_{22} = \frac{n^2}{2}-n\beta-1$,
$(\mathbf{M}_n)_{33} = -\frac{n^2}{2}-n\beta+1$,
$(\mathbf{M}_n)_{44} = -\frac{n^2}{2}+n\beta+1$, and all off‑diagonal entries are \((\mathbf{M}_n)_{ij}= -1\) for \(i=1,2\) (\(j\neq i\)) 
and \((\mathbf{M}_n)_{ij}= +1\) for \(i=3,4\) (\(j\neq i\)).

The MI gain $G$ for the $n$-th harmonic is defined as the maximum 
imaginary part of the eigenvalues $\lambda$ of the stability matrix $\mathbf{M}_n$ given 
above:
\begin{equation}
G(n, \beta) = \max\,\mathrm{Im}\bigl[\lambda(\mathbf{M}_n)\bigr].
\label{eq:gain_def}
\end{equation}
A positive $G$ signals exponential growth of the corresponding sideband pair $\pm n$, 
while $G=0$ indicates linear stability. For the three lowest harmonics, the stability behavior is as follows:
\begin{itemize}
\item $n=\pm1$ (fundamental modulation, frequency $\omega=1$): unstable for all $\beta>0$.
\item $n=\pm3$ (third harmonic, frequency $3\omega=3$): unstable in the interval 
$0.5<\beta<1.5$.
\item $n=\pm2$ (second harmonic, frequency $2\omega=2$): unstable only for $|\beta|<1$; 
for $\beta\ge 1$ the gain vanishes identically --- it lies inside a stable gap.
\end{itemize}

\begin{figure}[tbp]
\centering
\includegraphics[height=145pt,width=201pt]{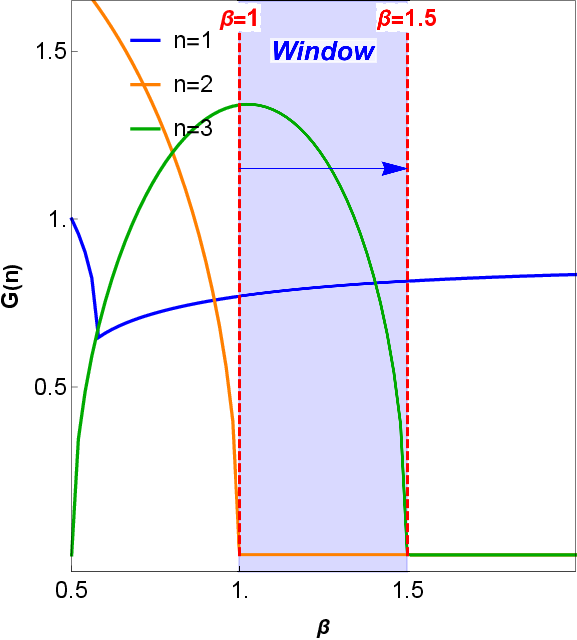}
\newline
\caption{Gain curves $G(n, \beta)$ of the first three modulation harmonics 
($n=1,2,3$) for the focusing Manakov system with $a=1$ and $\omega=1$. 
The second harmonic ($n=2$) becomes linearly stable (gain zero) for $\beta\ge 1$, 
while the first ($n=1$) and third ($n=3$) harmonics remain unstable in the 
shaded window $1\le\beta<1.5$. This window, where unstable modes alternate 
with a stable one, is the parameter range explored in this work. }
\label{fig:gain}
\end{figure}

Consequently, there exists a continuous window of the relative wavenumber,
\begin{equation}
\boxed{1 \le \beta < 1.5},
\label{eq:window}
\end{equation}
in which the first and third harmonics are modulationally unstable while the second 
harmonic remains strictly stable. This window is highlighted in FIG.~\ref{fig:gain}.

As shown above, we have identified a window $1\le\beta<1.5$ in which the $n=\pm1$ and $n=\pm3$ harmonics are modulationally unstable, while the $n=\pm2$ harmonic is linearly stable. In this window, a single pair of initial sidebands (e.g., $\omega$ and $3\omega$) can excite a higher-order MI process that involves multiple harmonics simultaneously. To capture the nonlinear interplay between the unstable modes and the passive generation of the $2\omega$ component, a solution containing three elementary breathers is required: two with frequency $n=1$ (the fundamental modulation admits a pair of non-degenerate breathers) and one with frequency $n=3$ (which supports a single breather). This constitutes a third-order AB of the Manakov system.

The general $M$-th order vector AB solution of the Manakov system~\eqref{eq:Manakov} 
can be constructed via the Darboux transformation~\cite{appl8}. For our purpose we take $M=3$ and 
select the three constituent eigenvalues as
\begin{equation}
\{\chi_1,\chi_2,\chi_3\} = \{\chi_+(\omega=1),\; \chi_-(\omega=1),\; \chi_+(3\omega=3)\},
\label{eq:chi_selection}
\end{equation}
where the first two correspond to the two distinct breather solutions at $n=1$, 
and the third to the single breather at $n=3$. 
Here $\chi_\pm(\omega)$ is given by
\begin{equation}
\chi_\pm(\omega) = 
\pm\sqrt{\beta^2-1+\frac{\omega^2}{4}-\sqrt{1+(\omega^2-4)\beta^2}}-\frac{\omega}{2}.
\label{eq:chi_omega}
\end{equation}

The resulting wave field of the $j$-th component $(j=1,2)$ can be written in the 
compact determinantal form
\begin{equation}
\psi^{(j)}[3](x,t) = \psi_0^{(j)}(x,t)\,
\frac{\det\mathcal{G}^{(j)}}{\det\mathcal{G}},
\label{eq:3rd_AB}
\end{equation}
where $\psi_0^{(j)}$ is the CW background~\eqref{eq:CW}. The $3\times3$ matrices 
$\mathcal{G}$ and $\mathcal{G}^{(j)}$ are built from the eigenfunctions associated 
with the three eigenvalues $\{\chi_m\}$ and their modulation frequencies 
$\{\omega_m\}=\{1,1,3\}$. Their explicit expressions are given in Sec.~I of the Supplemental 
Material.

In the following we are interested in the discrete Fourier spectrum of the breather, defined as
\begin{equation}
A_n^{(j)}(t) = \frac{1}{L}\int_{-L/2}^{L/2} \bigl[\psi^{(j)}(x,t)/\psi_0^{(j)}(x,t)-1\bigr]\,e^{-in x}\,dx,
\label{eq:spectrum_def}
\end{equation}
where the integration is over one period $L$ containing an integer number of fundamental wavelengths. While the early-time growth is dominated by the linearly unstable harmonics $n=\pm1,\pm3$, nonlinear coupling continuously generates all integer orders, most notably $n=\pm2$. Near the peak of the modulation cycle this passively driven component can become dominant.

To quantify the role of the stable harmonic, we define the dominance $\mathcal{D}^{(j)}$ at the temporal peak of the higher-order AB as
\begin{equation}
\mathcal{D}^{(j)}(\beta) = 
\frac{\max\bigl(|A_{2}^{(j)}(t_{\rm peak})|,\, |A_{-2}^{(j)}(t_{\rm peak})|\bigr)}
{\max_n\, |A_{n}^{(j)}(t_{\rm peak})|},
\label{eq:dominance}
\end{equation}
where $t_{\rm peak}$ is the time at which $|\psi^{(j)}(x,t)|$ reaches its global maximum. By definition, $0\le\mathcal{D}^{(j)}\le 1$. When $\mathcal{D}^{(j)}\approx 1$, the linearly stable $2\omega$ component is the dominant spectral feature at the breather's peak, i.e., the waveform is effectively controlled by a frequency inside the stability gap.

\begin{figure}[tbp]
\centering
\includegraphics[width=\columnwidth]{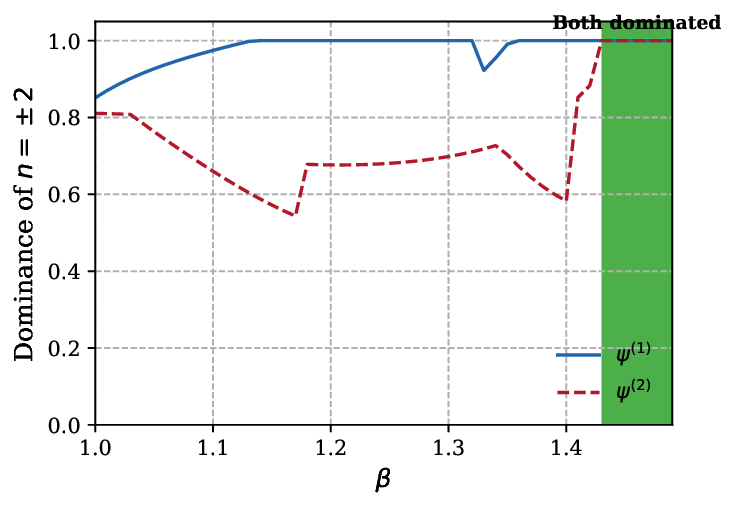}
\newline
\caption{Dominance $\mathcal{D}^{(j)}$ of the $n=\pm2$ sideband at the peak of the 
third-order Akhmediev breather as a function of $\beta$. 
The blue solid (red dashed) line represents the $\psi^{(1)}$ ($\psi^{(2)}$) component. 
The green shaded area marks the parameter interval where both polarizations are 
simultaneously dominated by the linearly stable $2\omega$ harmonic ($\mathcal{D}\approx1$). }
\label{fig:dom}
\end{figure}

We have computed $\mathcal{D}^{(1)}(\beta)$ and $\mathcal{D}^{(2)}(\beta)$ by scanning $\beta$ across the window $1\le\beta<1.5$ with a step of $0.01$, using the exact solution~\eqref{eq:3rd_AB}. The results are shown in FIG.~\ref{fig:dom}. Both components exhibit a clear transition from being dominated by the unstable harmonics ($n=\pm1,\pm3$) to being fully controlled by $n=\pm2$. For $\beta\gtrsim 1.40$, the dominance curves saturate at $\mathcal{D}^{(j)}\simeq 1$, signaling that the passively driven $2\omega$ breather has become the primary structure.
The critical value at which both components simultaneously reach $\mathcal{D}^{(j)}\approx 1$ is found to be $\beta_c\approx 1.43$. Therefore, the interval
\[
\boxed{\beta_c \le \beta < 1.5}
\]
defines the sub-window where a linearly stable frequency completely controls the peak dynamics of the higher-order AB.

\begin{figure}[tbp]
\centering
\includegraphics[height=115pt,width=261pt]{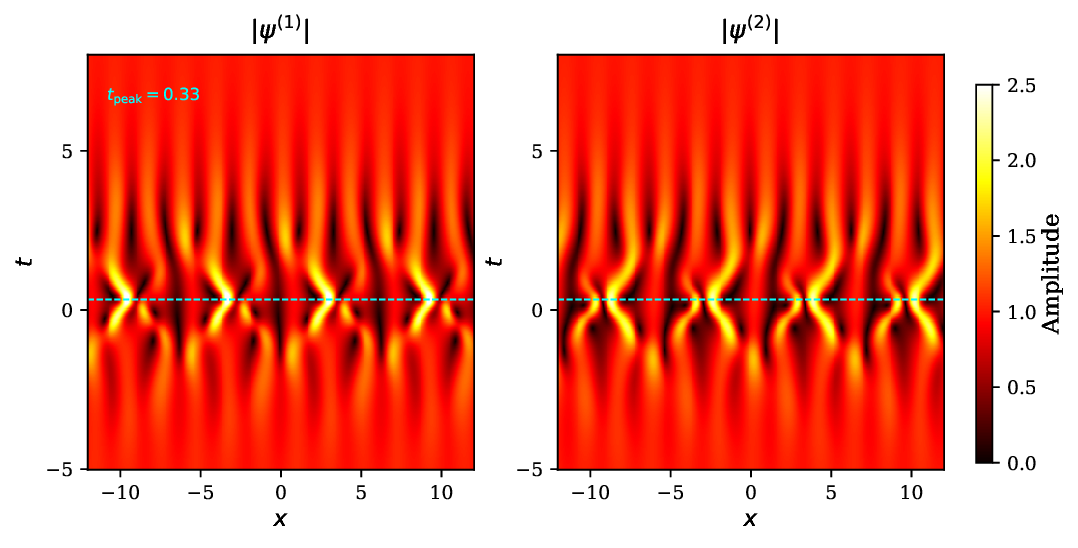}
\newline
\caption{Spatio-temporal evolution of $|\psi^{(1)}|$ (left) and $|\psi^{(2)}|$ (right) 
for the exact third-order Akhmediev breather at $\beta=1.44$. }
\label{fig:density}
\end{figure}

\begin{figure}[tbp]
\centering
\includegraphics[height=115pt,width=261pt]{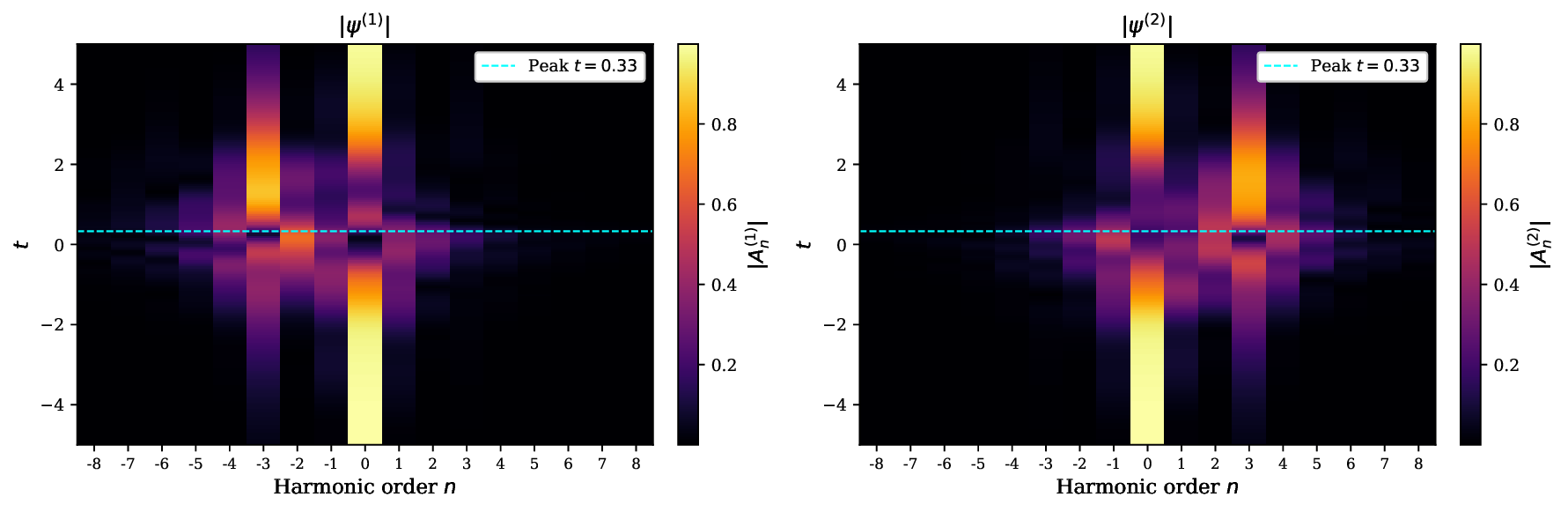}
\newline
\caption{Temporal evolution of the discrete Fourier amplitudes
$|A_n^{(1)}|$ (left) and $|A_n^{(2)}|$ (right) from $t=-5$ to $t=5$
at $\beta=1.44$. 
The horizontal dashed line marks the breather peak at $t=0.33$, 
where the linearly stable $n=\pm2$ harmonic becomes the dominant 
spectral component in both polarizations, exceeding the unstable 
modes $n=\pm1$ and $n=\pm3$ that seed the dynamics.}
\label{fig:spectrogramt}
\end{figure}

\begin{figure}[tbp]
\centering
\includegraphics[height=115pt,width=261pt]{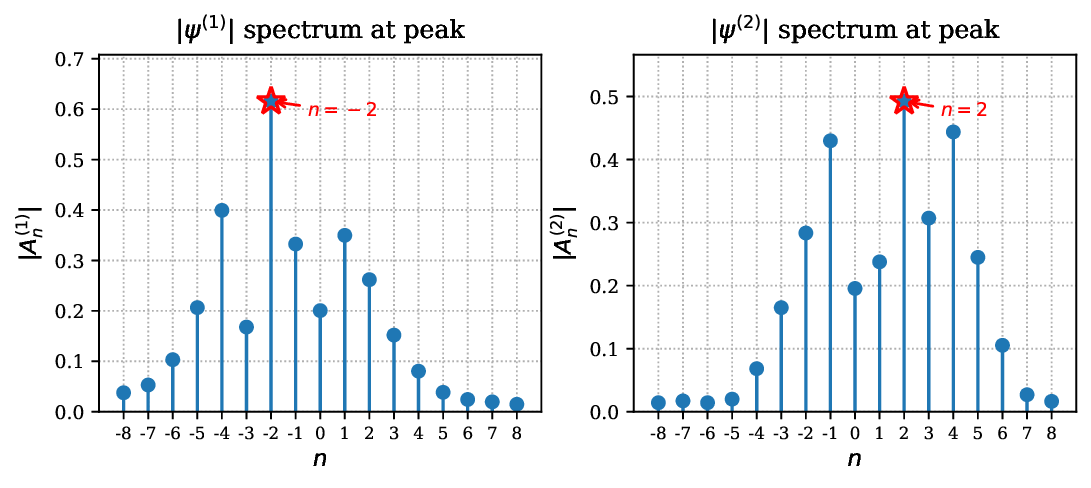}
\newline
\caption{Discrete Fourier amplitudes $|A_n^{(1)}|$ (left) and $|A_n^{(2)}|$ (right) 
at the peak of the third-order Akhmediev breather for $\beta=1.44$. 
In both polarization components, the $n=\pm2$ harmonic dominates the spectrum 
despite being linearly stable, exceeding even the unstable modes $n=\pm1$ and 
$n=\pm3$ that seed the breather. 
The red stars mark the maximum spectral component in each case. }
\label{fig:spectrum}
\end{figure}

To illustrate this mechanism concretely, we select $\beta=1.44$, which lies well inside 
the sub-window where both components are dominated by $n=\pm2$ at the peak. 
FIG.~\ref{fig:density} presents the spatio-temporal evolution of 
$|\psi^{(1)}|$ and $|\psi^{(2)}|$ obtained from the exact third-order AB solution.
This solution is, by construction, seeded by the two linearly unstable harmonics: 
the fundamental modulation $\omega=1$ and its third harmonic $3\omega=3$.
However,  the Fourier spectrum at the moment of maximum amplitude tells a different story. 
To further illuminate how the linearly stable $2\omega$ component is
passively amplified, we trace the full temporal evolution of the
discrete Fourier spectrum from the early linear stage to the nonlinear
peak. FIG.~\ref{fig:spectrogramt} displays the amplitudes of the
sidebands $|A_n^{(1)}|$ and $|A_n^{(2)}|$ as functions of time $t$
(from $-5$ to $5$) and harmonic order $n$ ($-8\le n\le 8$). 
At the earliest times only the linearly unstable modes $n=\pm1$
and $n=\pm3$ possess noticeable amplitudes. 
As the modulation grows, energy is continuously transferred into the
$2\omega$ band through four‑wave mixing, and the $n=\pm2$ sidebands
undergo a rapid increase that synchronizes with the growth of the
unstable harmonics. 
Near the breather peak, the $n=\pm2$ component becomes the largest spectral
feature in both polarizations, confirming that the waveform is
dominated by a frequency lying inside the stable gap.
FIG.~\ref{fig:spectrum} provides the central evidence of this work. Although the 
modulation originates exclusively from unstable harmonics, the largest Fourier amplitude 
at the breather peak belongs to the linearly stable $2\omega$ mode. The spectral maximum 
is therefore no longer determined by the MI gain profile but by nonlinear energy 
redistribution.

\begin{figure}[tbp]
\centering
\includegraphics[width=\columnwidth]{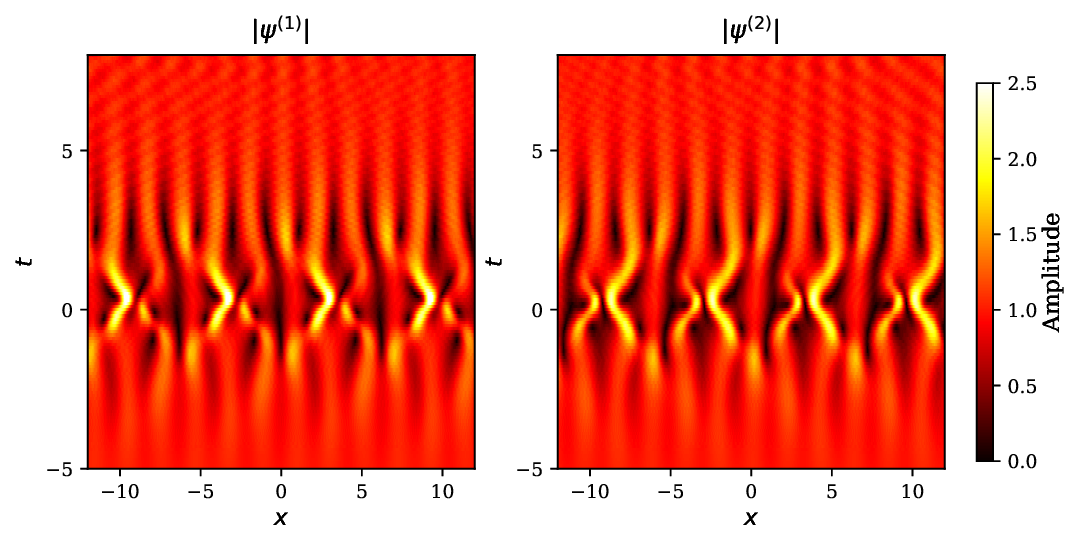}
\caption{Numerical evolution of $|\psi^{(1)}|$ and $|\psi^{(2)}|$ starting from the exact third-order AB profile at $t=-5$, using a split-step Fourier method. The dynamics closely follows the analytic solution of Fig.~\ref{fig:density}.}
\label{fig:numerical_evolution}
\end{figure}

\begin{figure}[tbp]
\centering
\includegraphics[width=\columnwidth]{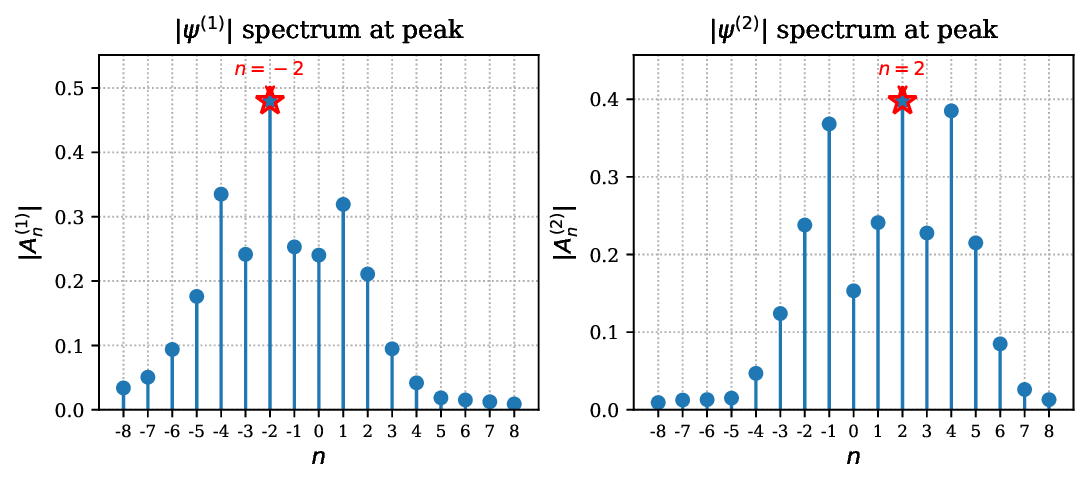}
\caption{Discrete spectra obtained from the numerical simulation at the peak times of each component. The quantitative agreement with the exact spectra of Fig.~\ref{fig:spectrum} confirms the robustness of the passive $2\omega$ amplification.}
\label{fig:numerical_spectra}
\end{figure}

To confirm this behavior, we have performed direct numerical integration of the Manakov system~\eqref{eq:Manakov} using a split-step Fourier method. The initial condition is taken as the exact third-order AB profile at $t=-5$. The numerical evolution (FIG.~\ref{fig:numerical_evolution}) reproduces the analytic prediction, and the spectra at the peak (FIG.~\ref{fig:numerical_spectra}) quantitatively match those of FIG.~\ref{fig:spectrum}, validating the robustness of the phenomenon.  In addition, to demonstrate that the $2\omega$ dominance does not rely on 
a fine‑tuned initial condition taken from the exact solution at $t=-5$, 
we have performed a separate numerical simulation starting from a 
CW background seeded only with the linearly unstable harmonics 
$\omega$ and $3\omega$ (i.e., without any initial $2\omega$ component). 
The precise form of this initial perturbation is provided in Sec.~II of the Supplemental 
Material. 
The resulting evolution (reported therein) confirms 
that the $2\omega$ sideband emerges spontaneously and becomes dominant 
at the breather peak without any initial seed, in agreement with 
the exact solution.

To prove that the amplification of the $2\omega$ sideband originates from
four-wave mixing, we isolate the nonlinear part of its dynamics.
Let $A_2^{(1)}(t)$ be the Fourier amplitude of the $n=2$ harmonic in the
first polarization component, extracted from the exact third-order AB
solution. The linearized evolution of the vector
$\mathbf{V}=(A_2^{(1)},A_2^{(2)},A_{-2}^{(1)*},A_{-2}^{(2)*})^{\!\mathrm{T}}$
is governed by the stability matrix $\mathbf{M}_2$ derived above.
The nonlinear contribution, i.e., the forcing term, is defined as
\begin{equation}
F_{\rm exact}(t) \equiv \frac{dA_2^{(1)}}{dt} - L_1(t),
\label{eq:F_exact}
\end{equation}
where $L_1(t)$ is the first component of the vector $-i\mathbf{M}_2\mathbf{V}(t)$.
The numerical derivative of $A_2^{(1)}$ is computed via a central-difference
gradient from the time series of the analytic solution. This construction
guarantees that $F_{\rm exact}(t)$ contains only nonlinear processes.

We model the forcing by the most resonant four-wave mixing channels that
can produce frequency $2\omega$. The leading terms are
$\omega+\omega\to2\omega$ and $3\omega-\omega\to2\omega$, together with
their frequency-reversed and cross-polarization counterparts.
Denoting the Fourier amplitudes of the first and second polarization
components by $A_n$ and $B_n$, respectively, we write
\begin{equation}
F_{\rm fit}(t) = \sum_{i=1}^{N_c} c_i\, \mathcal{M}_i(t),
\label{eq:F_fit}
\end{equation}
where $\{\mathcal{M}_i\}$ is a set of $N_c=20$ quadratic products of the
forms $A_m A_n$, $A_m A_n^*$, $B_m B_n$, $B_m B_n^*$, and the mixed terms
$A_m B_n$, $A_m B_n^*$, all satisfying the phase-matching conditions
$m+n=2$ or $m-n=2$ with $m,n\in\{\pm1,\pm3\}$.
The explicit list of channels is provided in Sec.~III of the Supplemental Material.
The complex coefficients $c_i$ are determined by a standard least-square
fit of $F_{\rm fit}(t)$ to $F_{\rm exact}(t)$ over the entire time
interval $t\in[-5,8]$.

\begin{figure}[tbp]
\centering
\includegraphics[width=\columnwidth]{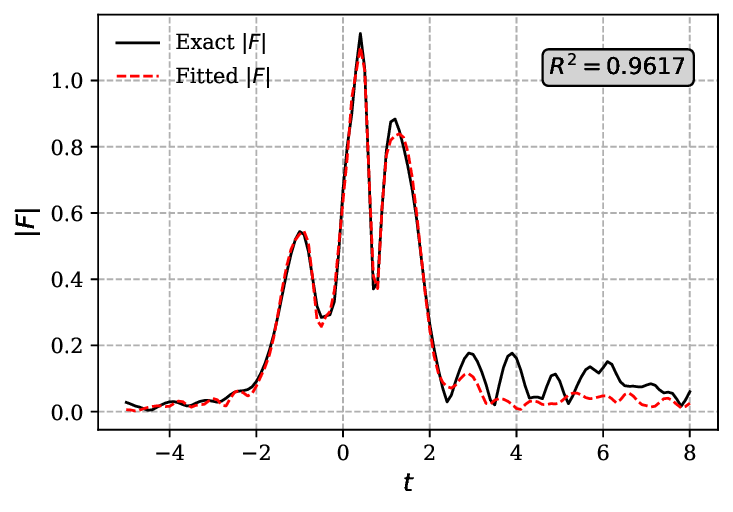}
\caption{Magnitude of the exact nonlinear forcing $|F_{\rm exact}(t)|$ (black solid line) 
and the fitted four-wave mixing model $|F_{\rm fit}(t)|$ (red dashed line) 
for the third-order Akhmediev breather at $\beta=1.44$. 
The forcing $F_{\rm exact}$ is extracted from the exact solution by subtracting 
the linear contribution given by the stability matrix $\mathbf{M}_2$. 
The fitted model $F_{\rm fit}$ is a linear combination of $20$ quadratic products 
of the unstable sideband amplitudes, covering all lowest-order four-wave mixing 
channels that satisfy $m+n=2$ or $m-n=2$ ($m,n\in\{\pm1,\pm3\}$). 
The excellent agreement ($R^2=0.96$) provides quantitative proof that the 
passive amplification of the $2\omega$ component is driven entirely by 
four-wave mixing of the unstable harmonics.}
\label{fig:forcing}
\end{figure}

The quality of the fit is quantified by the coefficient of determination
\begin{equation}
R^2 = 1 - \frac{\sum_{i} |F_{\rm exact}(t_i) - F_{\rm fit}(t_i)|^2}
                {\sum_{i} |F_{\rm exact}(t_i) - \bar{F}_{\rm exact}|^2},
\label{eq:R2}
\end{equation}
where $t_i$ are the discrete time points and $\bar{F}_{\rm exact}$ is
the time-averaged mean of $F_{\rm exact}(t)$. $R^2$ measures the fraction
of the total variation in the true forcing that is captured by the
four-wave mixing model: $R^2=1$ corresponds to perfect agreement.
FIG.~\ref{fig:forcing} compares $|F_{\rm exact}(t)|$ and $|F_{\rm fit}(t)|$
and reports $R^2=0.96$. Thus, 96\% of the driving power is explained
by quadratic products of the unstable harmonics.
This analysis provides quantitative  proof that the passive amplification of the linearly stable $2\omega$ component
is driven exclusively by four-wave mixing of the unstable harmonics.
It firmly establishes that the $2\omega$ component, rather than being bypassed as in the frequency-jumping scenario, becomes the leading spectral feature through passive nonlinear driving

In conclusion, we have analytically and numerically constructed an AB that is generated 
by unstable modes yet becomes spectrally dominated by a linearly stable harmonic at its peak. 
These results complement and extend the higher-order MI dynamics reported previously 
for the Manakov system~\cite{appl8}. Whereas higher-order MI was shown to bypass stable 
spectral gaps through frequency jumps, we demonstrate here that the same stable gap 
can instead host the dominant spectral component of an exact Akhmediev breather. 
Together, these two scenarios reveal that the nonlinear stage of MI is not constrained 
by the linear gain spectrum alone but is strongly reshaped by nonlinear spectral interactions.
\begin{acknowledgments}
The work of Sun is supported by the NSFC (Grants No. 12575001) and the Open Project of Key Laboratory of Mathematics and Information Networks (Beijing University of Posts and Telecommunications), Ministry of Education, China, under Grant No. KF202403. 
\end{acknowledgments}

\nocite{}

\end{document}